\documentclass[amsmath,amssymb,twocolumn]{revtex4}
\usepackage{dcolumn}
\usepackage{bm}
\usepackage{amsmath,mathrsfs}
\usepackage{graphicx,epsfig}
\usepackage{epsf}
\usepackage{color}

\begin{document}


\title{Wormhole solutions with a polynomial equation-of-state and minimal violation of the null energy condition }

\author{F. Parsaei}\email{fparsaei@gmail.com}
\author{S. Rastgoo}\email{rastgoo@sirjantech.ac.ir}
\affiliation{Physics Department , Sirjan University of Technology, Sirjan 78137, Iran}

\date{\today}


\begin{abstract}

This paper discusses wormholes supported by general equation-of-state  , resulting in a significant combination of
the linear equation-of-state and some other  models. Wormhole with a quadratic equation-of-state is studied as a particular example. It is shown  that the violation of null energy condition is restricted to some regions in the vicinity of the throat. The combination of barotropic and polytropic equation-of-state has been studied. We consider  fluid near the wormhole throat  in an exotic  regime which at some $r=r_{1}$, the exotic regime is connected to a distribution of asymptotically dark energy regime with $-1<\omega<-1/3$. We have presented wormhole solutions with small amount of exotic matter. We have shown that using different forms of equation-of-state has a considerable effect on the minimizing violation of the null energy condition. The effect of many parameters such as redshift as detected by a distant observer and energy density at the throat on the $r_1$ is investigated. The solutions are asymptotically flat and  compatible with presently available observational data at the large cosmic scale.


\end{abstract}

\maketitle

\section{Introduction}

 The significant development in wormhole physics has been started by Morris and Thorne \cite{WH}. Wormhole is an exact solution of the Einstein field equations which can be used as a hypothetical shortcut between points in a universe or between two different universes. Wormholes are not famous as black holes, but the possibility that wormholes and black holes are in fact very similar object has been presented\cite{black}. Wormholes have not been observed experimentally but there is not any observational reason to rule out this theory. Wormhole is a good candidate which would provide us with a practically unlimited possibility for interstellar travel.
  In wormhole study, a description of the matter content of the wormhole is essential to present wormhole physics. Morris-Thorne wormholes violate classical energy conditions \cite{Visser}.
Null energy condition (NEC) specified by $T_{\mu\nu}k^{\mu}k^{\nu}\geq0$, in which $k^{\mu}$ is any null vector and $T_{\mu\nu}$ stress-energy tensor.   The matter which violates the NEC  is called exotic. Exotic matter is the main ingredient in wormhole theory. It was shown that violation of the NEC is a generic and universal feature of the traversable wormhole in General Relativity (GR)\cite{Visser2}. Of course, wormholes with NUT parameter \cite{NUT} may be considered as traversable wormhole without exotic matter but, one should keep in mind that cosmological constant is exotic in this realm.
 Some researchers try to solve the problem of exotic matter by investigating wormhole in modified theories of gravity. Wormhole in modified gravitational theories such as Brans-Dicke \cite{Dicke}, $f(R)$ gravity \cite{tahereh}, curvature matter coupling \cite{modgravity2b}, and brane-world \cite{brane1,brane} has been studied. For a review on traversable wormhole solutions in modified theories of gravity, see \cite{modify} and references therein.  In most of these theories, an effective stress-energy tensor which contains the higher-order curvature has  appeared in the right side of the Einstein field  equations. So the effective stress-energy tensor violates NEC instead of ordinary matter.
Investigation of  wormholes in higher-dimensional spacetime was done \cite{ndimen}. Wormholes in the framework of   Lovelock\cite{Lovelock} which is considered as the most theory of gravitation in $n$ dimensions have been studied. Arkani-Hamed \textit{et al.} have studied Euclidean wormholes in string theory \cite{Arkani}.

Application of quantum inequalities imposed a bound on exotic energy densities \cite{equ}. The quantum effects allow local violation of energy conditions, but this bound implies that any inertial observer in flat spacetime cannot see an arbitrarily large negative energy density which lasts for an arbitrarily long period of time \cite{Kuh}. Kuhfittig has shown that  a wormhole supported by only small amounts of the exotic matter really can be traversable \cite{Kuh}.
 Wormholes with small amount of violation of the energy conditions in the framework of GR have been studied in the literature.
Somebody have used the cut and paste method to minimize the violation of NEC \cite{visser,cut}. In the cut and paste method, the single manifold $M$ is the combination of two spherically symmetric spacetime $M_{int}$ and $M_{ext}$ which the interior space is related to the wormhole and the exterior one is usually the vacuum space. The junction conditions originated from Israel junction condition, should be considered in this surgery. Since the wormhole geometry is confined to a local spacetime in the cut and paste method, the violation of NEC is due to this part and can be arbitrarily small. We believe that thin shells are a mathematical abstraction, for physical reasons, it is better to minimize the usage of the cut and paste method.

Theoretical physicists have used the different equation-of-state (EoS) to investigate cosmological as well as astrophysical phenomena. The EoS of fluid  supporting wormhole geometry is an essential equation in studying wormhole theory. Recent observations indicate that the Universe is spatially flat with low matter density and expanding with acceleration \cite{1}. Fluid with an EoS, $p=\omega \rho $, and positive energy density  is the most considerable one to explain the evolution of the cosmos. The regime with $-1<\omega\leq 0$ is called dark energy and the special case with $\omega<-1$ is called phantom. Accelerated expansion of the Universe is reachable for fluid with $\omega\leq -\frac{1}{3}$.
 The phantom fluid is found to be compatible with most of classical test of cosmology such as CMB (cosmic microwave background), anisotropy, and mass power spectrum. Although the exotic nature of phantom energy leads to some unusual cosmological consequences such as big rip, phantom fluid is a good candidate to support wormhole solutions. It violates the NEC and is speculated to be a possible driving late time cosmic acceleration. Many asymptotically flat\cite{phantom3,foad1} or not asymptotically flat\cite{phantom1,phantom2}  wormhole solutions with phantom fluid source have been presented. Wormholes with other forms of fluid, like Chaplygin EoS \cite{Chap}, polytropic EoS \cite{Poly}, and modified Chaplygin EoS \cite{jamil2} have been studied in the literature. Jamill\textit{ et al}. have studied wormhole supported polytropic phantom energy which is a generalization of phantom energy and in some cases Chaplygin-gas models \cite{mobasher}. In all of these papers, the violation of the NEC is inevitable in the whole of wormhole spacetime.

 Some authors have studied the wormhole with a variable EoS parameter in which $\omega=\frac{p}{\rho}$ is a function of radial coordinate\cite{rah,cat,lopez,Azr,Remo,foad2}. In some of these studies, the violation of energy condition is inevitable in the whole of spacetime \cite{rah}. In \cite{cat}, the cut and paste is used to find wormhole solutions of the finite size which minimize the violation of energy conditions. Lopez et al. have  assumed an EoS in which the sum of the energy density and radial pressure is proportional to a constant with a value smaller than that of the inverse area characterizing the system \cite{lopez}. They have found solutions which are not asymptotically flat.  Azreg-Aïnou by considering two barotropic equations
of state for lateral and radial pressures has presented wormholes with no gluing effects \cite{Azr}.
  In another method, Remo and Lobo have proposed  solutions in which phantom fluid is concentrated in the neighborhood of the throat to ensure the flaring out condition \cite{Remo}. Some of their solutions are asymptotically flat. They have investigated  the possibility that these phantom wormholes be sustained by their own quantum fluctuations. In \cite{foad2}, a general formalism to find asymptotically flat wormholes with variable EoS parameter has been presented. Also, the physical difference between the cut and paste method and  the intrinsically asymptotically flat wormholes which minimize the NEC has been discussed.
  Note that, it is not a necessary condition that the EoS of wormhole or any astrophysical object should be the same as the EoS of the Universe. The only condition is that the EoS of wormhole should be fitted by EoS of the Universe in the large cosmological scale. This motivation helps us to find wormhole solutions with EoS that is asymptotically linear but have additional terms. So this work is focussed on the exploration of wormhole solutions using the different form of EoS. We will seek solutions which restrict the violation of the NEC to some regions in the vicinity of the throat. The theory is GR but the matter content of wormhole is investigated in more details.
 Actually, we considered an EoS which has a linear term and some extra terms. This method can be used to sustain wormhole solutions with minimum violation of energy conditions. We study some new wormhole solutions which are asymptotically flat. We will present  solutions by considering a special shape function.

  The paper is organized as follows: In the next section,  the basic structure of wormhole theory is presented. In Sec \ref{sec3}, by considering a power-law shape function, wormhole with quadratic EoS is studied and solutions with minimum violation of the NEC are presented. Wormholes with a combination of barotropic and polytropic EoS are investigated in Sec \ref{sec4}. Discussions and  Concluding remarks are presented in the last section.


\section{The general model of wormhole theory}

Usually, the metric of a static and spherically symmetric wormhole is given by  \cite{WH}
\begin{equation}\label{1}
ds^2=-e^{2\phi(r)}dt^2+\left[ 1-\frac{b(r)}{r} \right]^{-1} dr^2+r^2\,d\Omega^2 \,,
\end{equation}
where $d\Omega^2= (d\theta^2+\sin^2\theta d\phi^2)$. At a minimum radial coordinate, $r_0$, with $b(r_0)=r_0$ the wormhole connects two different worlds or two distant parts of the same universe. Here $b(r)$ is called the shape or form function and $r_o$ is the throat of the wormhole. The function $\phi(r)$ is called redshift function because the gravitational redshift as measured by a distant observer is
\begin{equation}\label{f01}
z=\frac{\delta \lambda}{\lambda}=1-\frac{\lambda (r\rightarrow \infty
)}{\lambda(r=r_0)}=\frac{1}{\exp( \phi (r_0)) }.
\end{equation}
The redshift function should be finite everywhere to avoid the existence of the horizon. The conditions
\begin{equation}\label{f1}
\frac{(b-b' r)}{2b^2} > 0
\end{equation}
and
\begin{equation}\label{s1}
(1-b/r)>0.
\end{equation}
are necessary to support a traversable wormhole.  
Note that the prime denotes the derivative $\frac{d}{dr}$. In this paper, we investigate asymptotically flat wormhole solutions so the condition
\begin{equation}\label{f2}
\lim_{r\rightarrow \infty}\frac{b(r)}{r}=0 , \quad \lim_{r\rightarrow \infty}\phi(r)=0.
\end{equation}
is also imposed.
Considering an anisotropic fluid in the form $T^\mu_\nu={\rm diag}(-\rho, p,p_t,p_t )$ and using the  Einstein field equations, the following distribution of matter (with $8\pi G=c=1$) are obtained,
\begin{eqnarray}\label{2}
\rho&=&\frac{b'}{ r^2}, \\
\label{6}
p&=&\frac{2(r-b) \phi'}{ r^2}-\frac{b}{ r^3},\\
p_t&=&p+\frac{r}{2}\left[ p' + (\rho + p)\phi'  \right]\,
\label{pt}
\end{eqnarray}
where $\rho(r)$ is the energy density, $p(r)$ is the radial pressure and $p_t(r)$ is the lateral pressure.
Equation (\ref{pt}) can be considered as a consequence of the conservation of the stress-energy tensor, $T^{\mu\nu}{}_{;\mu}=0$. 
Although the wormhole solutions are anisotropic in general form, we can use an EoS for radial pressure and find lateral pressure through the Einstein equations. This is motivated by the discussion of
the inhomogeneities that may arise due to gravitational instabilities. Indeed, as the dark energy EoS represents a spatially homogeneous cosmic fluid and is assumed
not to cluster. It is also possible that inhomogeneities may arise due to gravitational instabilities. Thus anisotropic wormholes may have originated from density fluctuations in
the cosmological background, resulting in nucleation through the respective density perturbations. One
can also deduce the possibility that these structures are sustained by their own quantum fluctuations \cite{Remo}. Because of the aforementioned reason the pressure in the EoS may be regarded as a
radial pressure, and the tangential pressure may be deduced through the Einstein field equations. This method has been used extensively in the literature i.e. phantom wormholes \cite{phantom3,foad1,phantom1,phantom2}. Also Sushkov and  Kim have studied a time-dependent solution describing a spherically symmetric wormhole in a cosmological setting with a ghost scalar
field \cite{Sushkov}. More specifically, they have  shown that the radial pressure is negative throughout the spacetime, and
for large values of the radial coordinate, equals the lateral pressure, which demonstrates that the ghost scalar field
behaves essentially as dark energy.

There are many algorithms to construct wormhole geometry theoretically.  Assuming desired forms of the red shift and shape functions and then the corresponding matter field is determined by the Einstein field equations. In another method, the red shift or shape functions are obtained as solutions by solving the Einstein field equations with prescribed matter field configuration. The energy density and pressure in the first method usually are not realistic.  In the second algorithm, one may consider an EoS which has suitable properties to describe the physical matter.  The second algorithm has been used by many authors in the literature to find a wormhole solution with different forms of EoS \cite{phantom3,foad1,phantom1,phantom2,Chap,Poly,jamil2,mobasher,rah,cat}. In the present article, we used the same algorithm to find  asymptotically flat wormhole exact solutions with some other forms of EoS.

Generally, an EoS is added to  set of equations and then some strategies have been used to solve the equations and find  unknown functions. The EoS is very important because it explains the physical fluid which is essential to sustain the wormhole. Commonly, the essential fluid need to construct the wormhole geometry is not isotropic. We can use the radial pressure in the  EoS, $p=f(\rho)$, which was first presented in the study of phantom wormhole solutions \cite{phantom1}.
In recent studies, fluid  with a linear EoS  manifests itself as the source causing a rapid accelerated expansion of
the Universe on a large cosmic scale and EoS
\begin{equation}\label{n13}
 p=\omega \rho
 \end{equation}
 is the most famous in studying cosmos. As it was mentioned the accelerated expansion of the universe is reachable for fluid with $\omega\leq-1/3$. Fluid with $-1<\omega\leq-1/3$ and $ \rho>0$  satisfies the NEC which can cause accelerated expansion of the Universe. Phantom fluid violates the NEC every where in the spacetime but a linear EoS with $\omega>-1$  and an extra term may violate the NEC only in some region of the spacetime. So we try to find some solutions with barotropic EoS and a mixed energy density.

 We consider an EoS as follows:
\begin{equation}\label{f3}
p=\omega\rho(r)+g(\rho).
\end{equation}
 At large distance from the throat, the physics of wormhole should be compatible with the cosmos. So the condition
\begin{equation}\label{f4}
\lim_{\rho\rightarrow 0}\frac{g(\rho)}{\rho}=0,
\end{equation}
leads to an asymptotically linear EoS on the large scale.
Wormhole models building in cosmology based on  two main ingredients:
a theory of gravity and a description of the matter content of the universe. In this paper, the
gravity sector of the theory is not completely fixed. There are some free parameters. The matter sector is represented in the field equations by the energy-momentum tensor and some EoS rather than conventional ones have been considered. The conventional EoSs like linear, polytropic, Chaplygin, power-law and so on are models to describe cosmology. The wormhole theory has been explored in those models to achieve a good understanding of our universe. Since there are not any observational data about wormhole, it seems that the EoS of wormhole hasn't  a specific form. So, any EoS that has the necessary conditions in the local view and is compatible with cosmological known EoS in global view, can be considered as  a candidate.
 It is clear that wormhole geometries admit an EoS in the general form $p=f(\rho)$. Mathematically, this EoS can be expressed in a Taylor series so any general  EoS of the wormhole can be expressed in the form of Eq.(\ref{f3}).

The EoS (\ref{f3}) helps us to investigate a large class of wormhole solutions which violates the NEC only in some regions near the wormhole throat. Phantom wormholes violate the NEC in the whole of spacetime but the additional term $g(\rho)$ in Eq.(\ref{f3}) provides a way to construct wormhole solutions with the minimum violation of energy condition and considering $\omega>-1$ instead of phantom EoS in large scale.
In asymptotically flat spacetime at large radial coordinate, $\rho(r)$ and $p(r)$ should tend to zero. So to have compatibility with cosmos in large scale, we have
 \begin{equation}\label{9a}
 \lim_{\rho\rightarrow 0}p(\rho)=0.
\end{equation}
In the following, we consider some strategy to find exact wormhole solutions with different forms of $g(\rho)$.


\section{Wormhole with a quadratic equation of state }\label{sec3}

Ananda and Bruni have discussed cosmological dynamics and dark energy with a quadratic model of EoS \cite{Ananda}. They have shown this model evolves from a phantom phase asymptotically approaching a de Sitter phase instead of evolving to a big rip.
In \cite{Ananda} and references therein, it has been addressed the motivations behind considering a quadratic EoS. For example, the loop quantum gravity corrections which result in a modified Friedmann equation provides a quadratic EoS.  The modification is appearing as a negative term which is quadratic in the energy density. Also the studies of k-essence fields as unified dark matter (UDM) models indicated that a fluid with a closed-form barotropic EoS is a good candidate to describe the general k-essence field.
Now, we consider the same EoS  in the following form
   \begin{equation}\label{13}
 p(\rho)=\omega \rho+\omega_1\rho^2
 \end{equation}
It should be noted that the correct form of EoS is as follows
  \begin{equation}\label{13a}
 p(\rho)=\omega \rho+\frac{\omega_1}{\rho_c}\rho^2
 \end{equation}
but for the sake of simplicity, we set $\rho_c =1$.
This EoS is suitable to describe a standard fluid at high pressure and high energy density to vanishing presumer at low energy density($\lim_{\rho\rightarrow 0}p(\rho)=0$). Wormholes with a quadratic EoS has been studied by Rahaman \textit{et al}.\cite{Rahman}. They have considered the shape function as a series and  have found the form of coefficients of the series solution. They have studied the possibility of finding solutions for this particular shape function. They have not reached in a reasonable proof for the condition (\ref{s1}) in their research. They only discussed  some general mathematical possibility.
Let us study wormhole with a vanishing redshift function, putting $\phi=0$ in Eq.(\ref{6}), leads to
\begin{equation}\label{shape10}
 p=-\frac{b(r)}{r^3}.
 \end{equation}
So by considering (\ref{13}), one can get
\begin{equation}\label{shape11}
 r^2\,\omega b'+\omega_{1} b'^{2}=r\,b.
 \end{equation}
Finding exact solution for this equation is very difficult. Thus we try to find solutions with nonconstant redshift function. For the sake of simplicity, we set $r_0=1$ in the recent part of this paper. Several authors have investigated wormhole by considering power-law shape function\cite{foad1,phantom2,phantom1,Chap,Poly,mobasher}.
We choose a special shape function in the following form \cite{foad1}
\begin{equation}\label{shape1}
 b(r)=A\,r^\alpha+(1-A)
 \end{equation}
The mass function related to this shape function is
\begin{equation}\label{9b}
m(r)\equiv \int_{r_0}^r 4\pi r^2\rho dr=\frac{A\,r^\alpha-A}{2}.
\end{equation}
The condition $1>\alpha$ should be imposed to satisfy asymptotically flat condition (\ref{f2}). It is clear that for $\alpha>0$ the mass is unbounded and for $0>\alpha$ mass is bounded. This shape function shows an energy density in the form
 \begin{equation}\label{shape113}
 \rho(r)=A \alpha\,r^{(\alpha-3)}
 \end{equation}
which is a positive smooth function of $r$ for $A\,\alpha>0$ . This energy density function has a maximum at the throat
 \begin{equation}\label{shape113a}
 \rho_0=\rho(r_0)=A\, \alpha
 \end{equation}
  and tends to zero at large infinity distance from the throat. Equation (\ref{shape113a}) relates the constant $A$ and $\alpha$ to physical parameter $\rho_0$.
  Someone may believe that specifying a shape function lacks physical justification and motivation for the stress-energy tensor. But we can consider another alternative approach that has been used by many authors.  As it was done in \cite{foad1,phantom1}, one can consider a specific energy density profile instead of considering a known shape function. In this context, the energy density profile (\ref{shape113}) can be considered known which leads to the shape function (\ref{shape1}). We think that these two approaches are intrinsically the same.  So we consider only the known shape function instead of considering a smooth energy density distribution. The presented algorithm could be used for  other forms of energy density distribution.

Because of the aforementioned properties, the defined shape function is a good candidate to find wormhole exact solutions. Now, we try to find solutions for $\phi(r)$ by using the Einstein field equation. Using Eq.(\ref{6}) leads to
\begin{equation}\label{shape13}
 \phi(r)=\frac{1}{2}\int(\frac{b(r)+r^3\,p(r)}{r(r-b(r))})dr
 \end{equation}
  Putting $p(r)$ from (\ref{13}) and (\ref{shape113a}) in this equation yields
  \begin{equation}\label{shape14}
 \phi(r)=\frac{1}{2}\int \frac{(A+\omega\,A)r^\alpha +1-A+\omega_{1}A^2 \alpha^2 r^{2\alpha-3}}{r(r-A r^\alpha +A-1)}dr.
 \end{equation}
  \begin{figure}
\centering
  \includegraphics[width=3.2in]{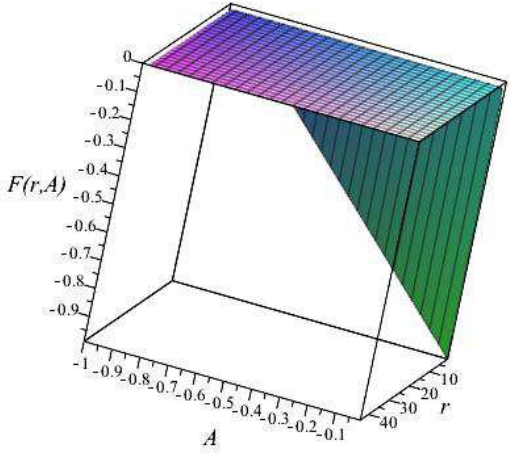}
\caption{The plot depicts the function $F(r,A)$ against $r$ and $A$. It is clear that $F(r,A)$ is negative thorough the entire range of $r$ and $A$ which means the NEC is violated everywhere.
 See the text for details.}
 \label{fig1}
\end{figure}

 Calculating this integral for a general form of $A$ and $\alpha$ is difficult,  so we analyzed some special cases.
 As the first example, consider $\alpha=-1$ which leads to
   \begin{eqnarray}\label{phi1}
 \phi(r)&=&\frac{c_1}{r^4}+\frac{c_2}{r^3}+\frac{c_3}{r^2}+\frac{c_4}{r} \nonumber  \\
 &+&c_5 \ln (r)+c_6 \ln (r-1)+c_7 \ln (r+A)
 \end{eqnarray}
with
\begin{eqnarray}\label{phi2}
c_1&=&\frac{\omega_1 A}{8},\,c_2=-\frac{\omega_1 }{6}(1-A),\,c_3=-\frac{\omega_1 }{4}(1-A-\frac{1}{A}),
   \nonumber  \\
 c_4&=&-\frac{\omega }{2}(1-A-\frac{1}{A}+\frac{1}{A^2}),
 \nonumber  \\
c_5&=&\frac{1 }{2}(\omega+\omega_1 (1-A-\frac{1}{A}+\frac{1}{A^2}-\frac{1}{A^3})+1),
 \nonumber  \\
c_6&=&\frac{1 }{2(A+1)}(1-\omega A+\omega_1 A^2),
 \nonumber  \\
 c_7&=&\frac{1 }{2(A+1)}(A-\omega +\frac{\omega_1}{A^3}).
\end{eqnarray}
To avoid from horizon, we should set $c_5=c_6=c_7=0$ which leads to
  \begin{equation}\label{phi3}
 \omega=\frac{A^6-1}{A^5-A},\qquad \omega_1=\frac{A^2-A^4}{A^4-1}.
 \end{equation}
Equation (\ref{shape113}) implies that for a positive energy density, $A$ should be negative. Defining the function
\begin{eqnarray}\label{phi4}
F(r,A)&=&\rho(r,A)+p(r,A)\nonumber  \\
&=&((\omega(A)+1)+\omega_1(A)\rho(r,A))\rho(r,A)
 \end{eqnarray}
help us to investigate the violation of the NEC. The function $F(r,A)$ is depicted as the surface in Fig. \ref{fig1}, which is negative through the spacetime so the NEC is violated everywhere.
Let us try to find solutions which satisfy the NEC in some regions of spacetimes. We consider $A=1$ and $\alpha=1/2$. Equation (\ref{shape14}) provides
 \begin{eqnarray}\label{shape15}
 \phi(r)&=&\frac{1}{2}(\frac{\omega_1}{2} (\frac{1}{r^{1/2}}+\frac{1}{2r}+\frac{1}{3r^{3/2}}+\frac{1}{4 r^2}+\frac{1}{5 r^{5/2}})
 \nonumber  \\
 &+&(\frac{\omega}{2}+1+\frac{\omega_1}{4})\ln(1-\frac{1}{\sqrt{r}})).
 \end{eqnarray}
  \begin{figure}
\centering
  \includegraphics[width=3.2in]{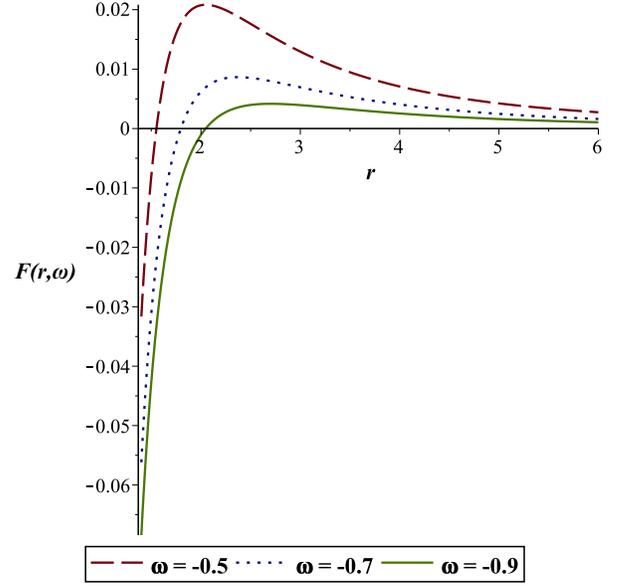}
\caption{ $F(r,\omega)$ for $\omega=-0.5$ (dashed line),$\omega=-0.7$ (dotted line), $\omega=-0.9$ (solid line), against $r$ . It is clear that $F(r,\omega)$ is positive throughout some range  $r>r_1$  which implies the NEC is satisfied. See the text for details.}
\label{fig2}
\end{figure}
The condition of an event-horizon-free spacetime
requires that $\phi(r)$ be finite everywhere. Therefore the coefficient of  $\ln(r)$  should be equal to zero. This yields
\begin{equation}\label{shape16}
 \omega_1=-(2\,\omega +4).
 \end{equation}
 Using Eqs.(\ref{f01}) and (\ref{shape15}), one can easily show that
 \begin{equation}\label{f10}
 \omega_1=-\frac{240}{137}\ln(z).
 \end{equation}
 It relates the redshift as detected by a distant observer and $\omega_1$.
 Since $-1<\omega<-1/3$ is related to dark energy regime, it is easy to show that
\begin{equation}\label{shape17}
 -\frac{10}{3}<\omega_{1}<-2,
 \end{equation}
 is acceptable in this regime. Now, we devote some words to the possibility of the violation of the NEC. We  define the function
\begin{eqnarray}\label{shape18}
F(r,\omega)&=&\rho(r)+p(r,\omega) \nonumber  \\
&=&((\omega+1)-2(\omega+2)\rho(r))\rho(r).
 \end{eqnarray}

\begin{figure}
\centering
  \includegraphics[width=3.2in]{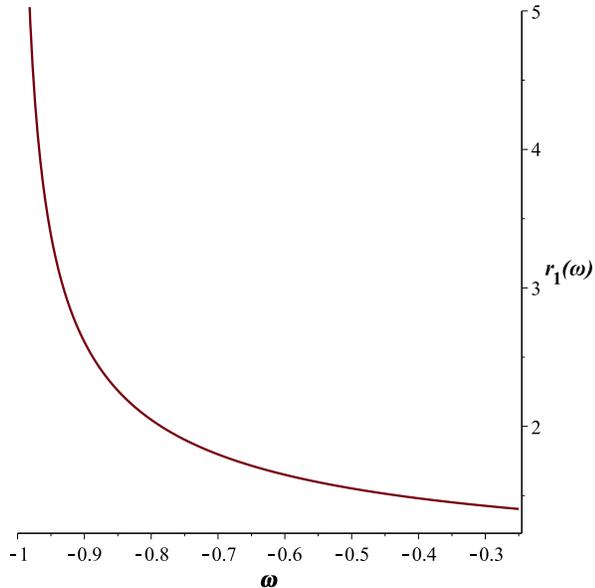}
\caption{$r_1$ as a function of $\omega$. It is transparent that $r_1$ decreases as  $\omega$ increases. See the text for details.}
\label{fig3}
\end{figure}
\begin{figure}
\centering
  \includegraphics[width=3.2in]{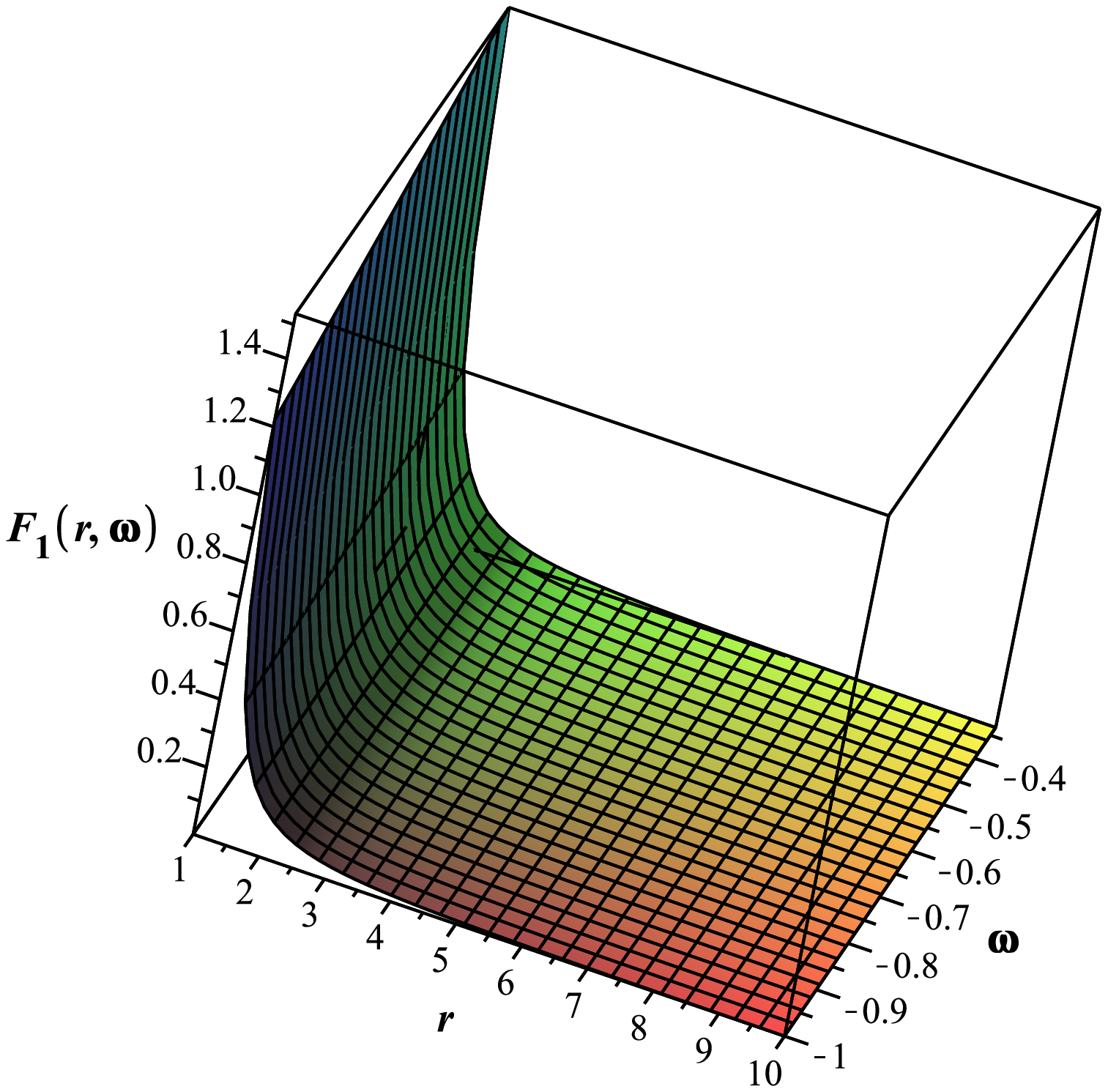}
\caption{The plot depicts the function $F_1(r,\omega)$ against $r$ and $\omega$. It is clear that $F_1(r,\omega)$ is possitive thorough the entire range of $r$ and $\omega$ which means the NEC is satisfied everywhere.
 See the text for details.}
 \label{fig22}
\end{figure}

The sign of this function indicates the violation of the NEC. The negative sign corresponds to exotic matter. We have plotted $F(r,\omega)$ against $r$ in Fig. \ref{fig2} for some different $\omega$. It is clear that the violation of the NEC is restricted to some region in the vicinity of the throat.
 Some explicit calculations yield
\begin{equation}\label{shape19}
r_1(\omega)=(\frac{\omega+2}{\omega+1})^{(\frac{2}{5})}.
 \end{equation}
  Here $r_1$ is the radial where the sign of $F(r,\omega)$ changes. We have plotted $r_1$ as a function of $\omega$ in Fig. \ref{fig3}. This figure shows that as $\omega$ increases $r_1$ decreases. Since $\omega<-1/3$ leads to accelerated expansion of the Universe, the minimum of $r_0$ is given by
 \begin{equation}\label{shape20}
r_{min}=\lim_{\omega \longrightarrow -\frac{1}{3}}(\frac{\omega+2}{\omega+1})^{(\frac{2}{5})}=(\frac{5}{2})^{\frac{2}{5}}.
 \end{equation}
  It is also obvious that $\lim_{\omega \longrightarrow -1}r_{1}(\omega)\longrightarrow \infty$.
In general, we can conclude that the selected shape function (\ref{shape1}) can be used to construct wormhole geometries with a  quadratic EoS fluid in which the violation of the NEC is considerable only near the wormhole throat. Although the violation of the NEC is resulted from the flaring out condition which appears directly in $\rho+p$, it is necessary  to check the other NEC $\rho+p_t\geq0$. So we have defined the function
\begin{equation}\label{shape20a}
F_1(r,\omega)=\rho(r)+p_t(r,\omega).
 \end{equation}
We have plotted $F_1(r,\omega)$ against $r$ and $\omega$ in Fig. \ref{fig22}. Since the $F_1(r,\omega)$ is positive through the entire range of $r$ and $\omega$, we can conclude that $\rho+p_t\geq0$ holds everywhere. 

  In the next sections, we will try to find solutions with a more general EoS. The class of solutions has been presented in this section seems to be a special case of the solutions in the next section.

\section{Combination of barotropic and polytropic  equation of state }\label{sec4}
Wormhole with a polytropic EoS has been studied in the literature \cite{Poly,mobasher}. On the other side, many papers have discussed wormhole with a barotropic EoS \cite{foad1,phantom1,phantom2,baro}. Now, we consider wormhole with a combination of barotropic and polytropic EoS :
\begin{equation}\label{N1}
p=\omega\rho+\omega_1\rho^n,
\end{equation}
An important feature of this EoS is the reduction of exotic matter in constructing wormhole theory.
We use the method  of the previous section to find exact wormhole solutions. If  we let $A = 1 $ in the shape function (\ref{shape1}) and use Eq. (\ref{N1}), the form of integral (\ref{shape14})  becomes as follows
\begin{equation}\label{shape22}
\phi(r)=\frac{1}{2}((1+\omega\alpha)\phi^{\alpha}_{1}+\omega_1\alpha^n \phi^{\alpha}_{m}).
 \end{equation}
where
\begin{eqnarray}\label{shape23}
\phi^{\alpha}_{1}&=&\int \frac{1}{r(r^{(1-\alpha)}-1)}dr=\frac{\ln(1-\frac{1}{r^{1-\alpha}})}{1-\alpha},
\nonumber \\
\phi^{\alpha}_{m}&=&\int \frac{1}{r^m(r^{(1-\alpha)}-1)}dr,
 \nonumber \\
 m&=&(1-n)(\alpha-3)+1.
 \end{eqnarray}
Considering the derivative of redshift function finite at the throat (the term behind integral (\ref{shape13} should be finite at the throat) leads to
\begin{equation}\label{shape24}
\omega_1=-\frac{1+\omega \alpha}{\alpha^n}.
 \end{equation}
The solutions of $\phi^{\alpha}_{m}$ for a general $\alpha$ and $m$ appear in the hypergeomtric function form, so we will solve  some special cases. An interesting solution corresponds to $\alpha=\frac{1}{2}$ and
 \begin{equation}\label{shape25}
m=\frac{s}{2}, \qquad s=3,4,5,... .
 \end{equation}
 \begin{figure}
\centering
  \includegraphics[width=3.2 in]{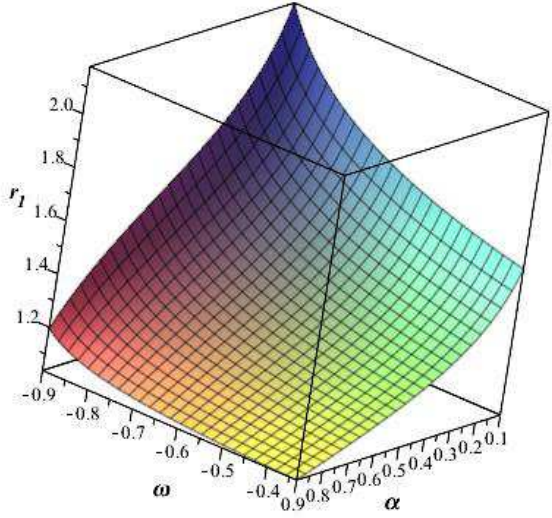}
\caption{The plot depicts the function $r_1(n,\omega,\alpha)$ as a function of $\omega$ and $\alpha$ for $n=3$. It is clear that  $r_1$ decreases as $\alpha$ or $\omega$ increases. See the text for details.}
\label{fig4}
\end{figure}
  For these values, we can verify that $n=\frac{3+s}{5}$ and
 \begin{equation}\label{shape26}
\phi^{\frac{1}{2}}_{m}=\phi^{\frac{1}{2}}_{1}+\sum^{2m-2}_{j=1}\frac{2}{j\,r^{j/2}}.
 \end{equation}
 Finally, from Eqs.(\ref{shape22}-\ref{shape26}) for $n=\frac{3+s}{5}$ and $\alpha=\frac{1}{2}$ one can get
  \begin{equation}\label{shape27}
\phi=-(1+\frac{\omega}{2})\sum^{5n-5}_{j=1}\frac{1}{j\,r^{j/2}}.
 \end{equation}
 By taking into account Eqs. (\ref{f01}) and (\ref{shape27}), one can verify that
 \begin{equation}\label{sha}
z=e^{((\omega/2+1)H_{5n-5})},
 \end{equation}
 where $H_{5n-5}$ is the $(5n-5)$-th harmonic number. Equation (\ref{sha})  shows the relation between $z$ and $n$.
  To check the NEC, we use the function
  \begin{equation}\label{shape28}
F(n,r,\omega)=\rho(r)+p(r,\omega)=((\omega+1)+\omega_1\rho(r)^{n-1})\rho(r).
 \end{equation}
 It is clear that $F(n,r,\omega)$ depends on $n,\omega$ and $r$. Since $\rho>0$, one can deduce that
    \begin{equation}\label{shape29}
r_1(n,\omega,\alpha)=(\alpha \frac{1+\omega}{1+\alpha \omega})^{\frac{1}{(n-1)(\alpha-3)}}.
 \end{equation}
 \begin{figure}
\centering
  \includegraphics[width=3 in]{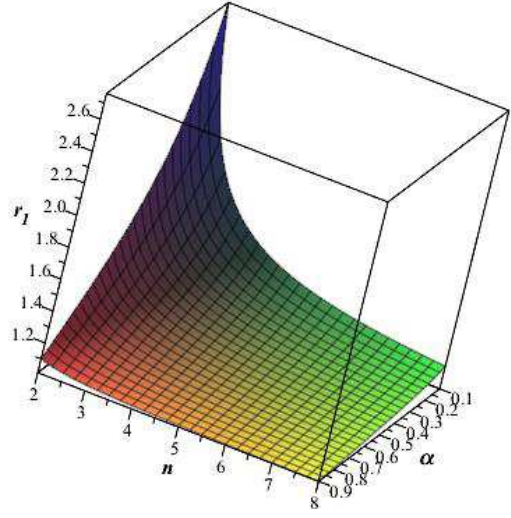}
\caption{The plot depicts the function $r_1(n,\omega,\alpha)$ as a function of $\alpha$ and $n$ for $\omega=-\frac{1}{2}$. It is clear that  $r_1$ increases as $\alpha$ or $n$ decreases. See the text for details.}
\label{fig5}
\end{figure}
 Here $r_1$ is the radial where the sign of $F(n,r,\omega)$ changes. We have plotted $r_1(n,\omega,\alpha)$ as a function of $\omega$ and $\alpha$ for $n=3$ in Fig. \ref{fig4}. This figure shows that the value of $r_1$ increases as $\alpha$ or $\omega$ decreases, so in the solutions with larger $\omega$ or $\alpha$, the violation of the NEC is limited to a smaller space near the throat. We can use the same analyzes for the dependence of $r(n,\omega,\alpha)$ on $n$ and $\alpha$ or $n$ and $\omega$ when the third parameter is fixed. Figs. \ref{fig5} and \ref{fig6} describe these dependencies. Figure \ref{fig5} indicates that $r_1$ is an ascending function as $\alpha$ or $n$ decreases. Also, Fig. \ref{fig6} explains that $r_1$ is an ascending function as $\omega$ or $n$ decreases which shows that all of three parameters generally have the same behaviour. Observe that as $n\longrightarrow\infty$ then $r_1\longrightarrow 1$ which means that for larger $n$ the violation of the NEC is limited to a very thin radial in the vicinity of the throat.

\begin{figure}
\centering
  \includegraphics[width=3 in]{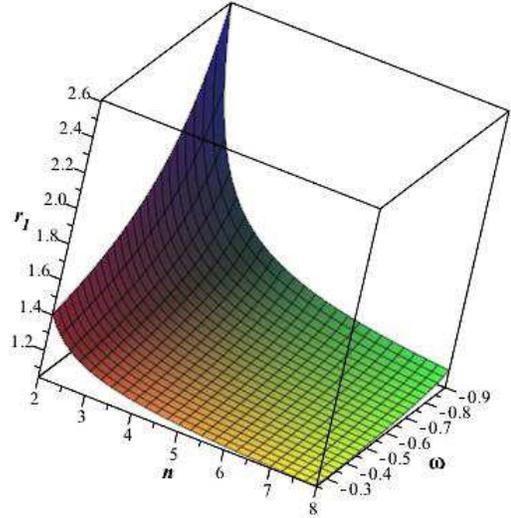}
\caption{The plot depicts the function $r_1(n,\omega,\alpha)$ as a function of $\omega$ and $n$ for $\alpha=-\frac{1}{2}$. It is clear that  $r_1$ increases as $\alpha$ or $n$ decreases. See the text for details. }
\label{fig6}
\end{figure}
\begin{figure}
\centering
  \includegraphics[width=3 in]{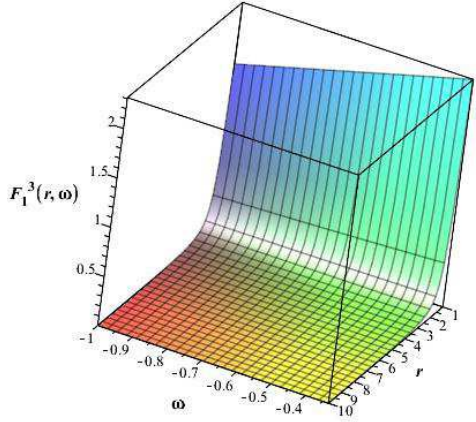}
\caption{The plot depicts the function $F^3_1(r,\omega)$ as a function of $\omega$ and $r$ for. It is clear that  $F^3_1(r,\omega)$ is positive through the entire range of $\omega$ and $r$. This shows that condition $\rho+p_t\geq0$ holds everywhere. See the text for details. }
\label{fig77}
\end{figure}
For this class of solutions, investigating the condition $\rho+p_t\geq0$, for a general $n$ is too complicated. So, we will check this condition for each $n$ individually. According to the previous section, we can define the function
\begin{equation}\label{shape20aaa}
F^n_1(r,\omega)=\rho(r)+p_t(r,\omega)
 \end{equation}
for each $n$ then we can plot $F^n_1(r,\omega$ against $r$ and $\omega$. As an example, we have plotted $F^3_1(r,\omega)$ against $r$ and $\omega$ in Fig.\ref{fig77}. This figure guarantees that $\rho+p_t\geq0$ holds everywhere. 
Let us seek a special example for wormhole solutions violating the NEC only in the vicinity of the wormhole throat in detail. If we set  $\alpha=\frac{1}{2}$, $\omega=-\frac{2}{3}$ and $n=3$ , leading to the line element
\begin{equation} \label{28a}
ds^2=- e^{-\frac{4}{3}\sum^{10}_{j=1}\frac{1}{jr^{j/2}}}dt^2
+\frac{dr^2}{1-\frac{1}{\sqrt{r}}}
+r^2\,d\Omega^2.
\end{equation}
The stress-energy tensor components are as follows
\begin{eqnarray}\label{29}
\rho(r)&=&\frac{1}{2\,r^{5/2}} ,\\
p(r)&=&-\frac{1}{3\,r^{5/2}}(1+\frac{2}{r^5})  ,\\
p_{t}(r)&=&\frac{1}{36\,r^{13}}(3r^{21/2}+r^{10}+r^{19/2}+r^9+r^{17/2}\nonumber \\
&+&r^8+r^{15/2}+r^7+r^{13/2}+r^6+133r^{11/2}\nonumber \\
&-&4r^5-4r^{9/2}-4r^4-4r^{7/2}-4r^3-4r^{5/2}\nonumber \\
&-&4r^2-4r^{3/2}-4r-4\sqrt{r}) .
\end{eqnarray}
Then $r_1=4^{1/5}\simeq1.32$ and the  total amount of exotic matter can be measured by\cite{zas}
\begin{equation}\label{f12}
I=8\pi \int_{r_0}^{r_1} (\rho+p)r^{2}dr .
\end{equation}
\begin{figure}
\centering
  \includegraphics[width=3 in]{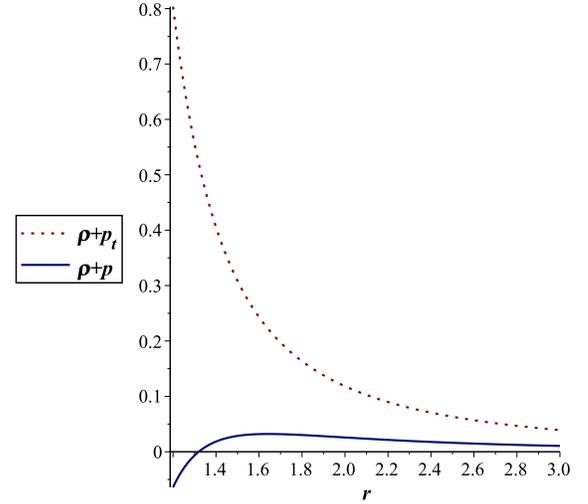}
\caption{The plot depicts the function $p+\rho$ (solid line) and $p_t+\rho$ (dotted line)  as a function of $r$ where $n=3, \omega=-2/3$ and $r_1\simeq 1.32$. It is clear that  for $r > r_1$ the NEC is satisfied. See the tex for details}
\label{fig7}
\end{figure}
We have plotted $\rho+p$ and $\rho+p_t$ as a function of $r$ in Fig. \ref{fig7} which implies that the NEC is violated only in the interval $r_0\leq r <r_{1} $ with the total amount of the NEC violation $I\simeq-8\pi\times 0.056$ and $z=e^{3/4 h_{10}}\simeq 10.58$. To summarize, the line element (\ref{28a}) is a special example of wormholes with an unbounded mass which has been constructed by a fluid in the form of Eq. (\ref{N1}). One can construct many different wormholes solutions by considering other forms of $b(r)$ and $p(\rho)$  with the minimum violation of the NEC  by carefully fine-tuning the parameters to find $\phi(r)$.

The same analyse can be used for a general EoS in the form
\begin{equation}\label{N2}
p=\omega\rho+\sum^{N}_{i=1}\omega_i\rho_i^{n_i}
\end{equation}
where $n_i>1$. The physical properties of this EoS when $n_i$ is an integer has been investigated in \cite{Vis}. The condition, avoiding horizon at the throat, gives
\begin{equation}\label{N3}
1+\omega \alpha+\sum^{N}_{i=1}\omega_i\alpha^{n_i}=0.
\end{equation}
It is easy to show that
\begin{equation}\label{N4}
\phi(r)=\frac{1}{2}((1+\omega\alpha)\phi_1^\alpha+ \sum^{N}_{i=1}\omega_i\alpha^{n_i}\phi^{\alpha}_{m_i})
\end{equation}
where
\begin{equation}\label{N5}
m_i=(1-n_i)(\alpha-3)+1.
\end{equation}
As it was mentioned before, calculating $\phi_{m_i}^{\alpha}$ for general $\alpha$ and $m_i$ leads to hypergeometrical functions. So one can study the special cases by choosing the suitable values for $\alpha$ and $m_i$. Again, we put $\alpha=\frac{1}{2}$ and $m_i=\frac{s}{2}, s=3,4,5,...$ or $n_i=\frac{3+s}{5}$, then
\begin{equation}\label{N6}
\phi(r)=(\sum^{N}_{i=1}\omega_i(\frac{1}{2})^{n_i}\sum^{5n_i-5}_{j=1}\frac{1}{j r^{\frac{j}{2}}}).
\end{equation}
For example,  by using Eq. (\ref{N6}) for the case where the only non vanishing $n_i$ are $n_1=2$ and $n_2=3$ with
\begin{equation}\label{N6a}
p(\rho)=\omega\rho+\omega_1\rho^2+\omega_2\rho^3,
\end{equation}
one can find
\begin{equation}\label{N6b}
\phi(r)=(\frac{\omega_1}{4}\sum^{5}_{j=1}\frac{1}{j r^{\frac{j}{2}}}+
(\frac{\omega_2}{8})\sum^{10}_{j=1}\frac{1}{j r^{\frac{j}{2}}}).
\end{equation}
then for $\omega=-\alpha=-\frac{1}{2}$, Eq. (\ref{N3}) yields
\begin{equation}\label{N7}
\omega_1 =-\frac{6+\omega_2}{2}.
\end{equation}
This implies that we can not choose all of  $\omega_i$ in Eq.(\ref{N3}) freely.
 It is easy to show that $p+\rho$ changes sign in
\begin{eqnarray}\label{N8}
\rho_\pm &=&\frac{-\omega_1\pm\sqrt{\omega^2_1-4\omega_2(1+\omega)}}{2\omega_2}\nonumber  \\
&=&\frac{3+\frac{\omega_2}{2}\pm\sqrt{(3+\frac{\omega_2}{2})^2-2\omega_2}}{2\omega_2}.
\end{eqnarray}

\begin{figure}
\centering
  \includegraphics[width=3 in]{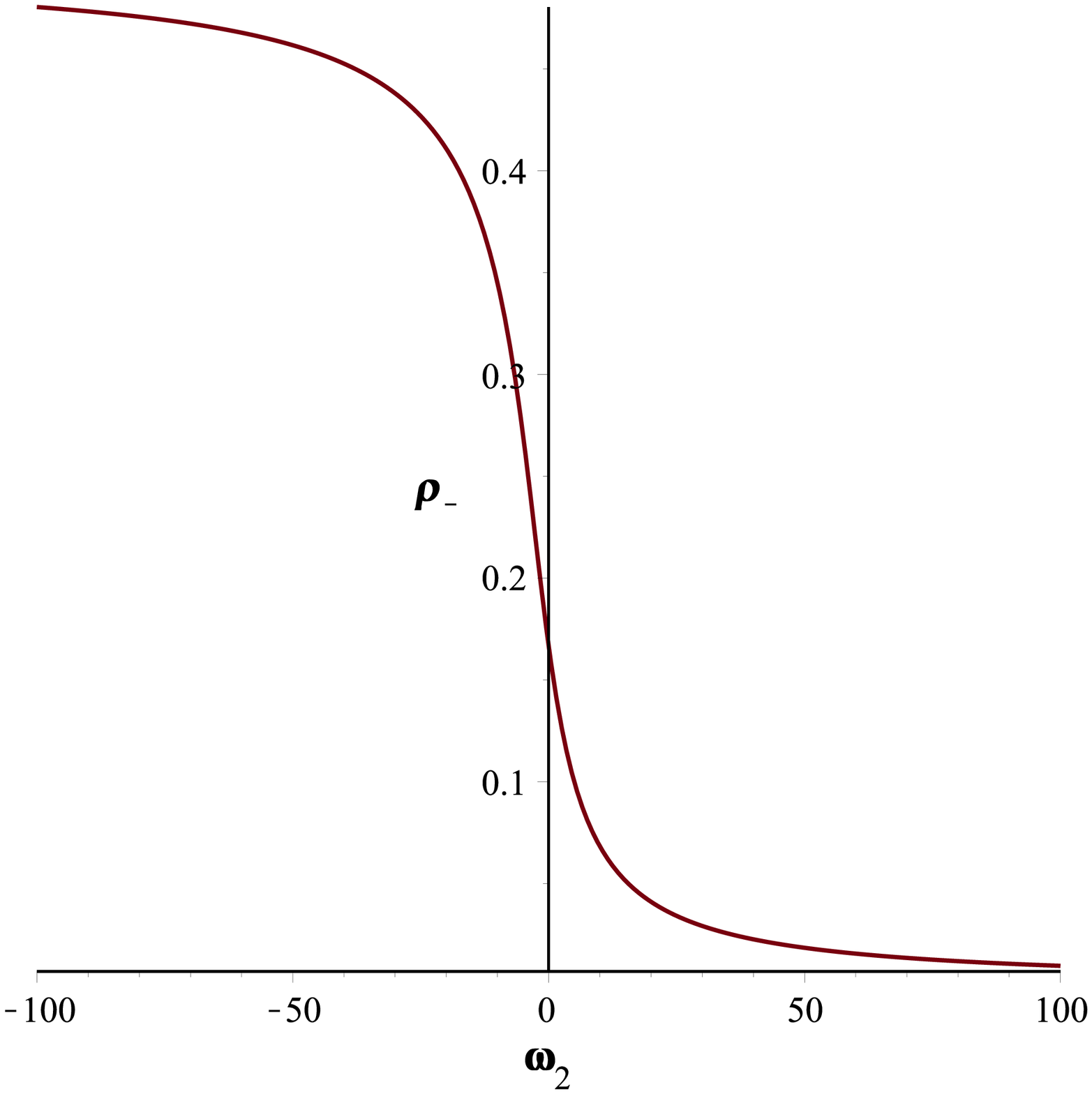}
\caption{The plot depicts the function $\rho_-(\omega_2)$ as a function of $\omega_2$ . It is clear that  $\rho_-$ is  in the possible range $0\leq \rho\leq \frac{1}{2}$ so $\rho_-$ is  acceptable. See the tex for details.}
\label{fig8}
\end{figure}
\begin{figure}
\centering
  \includegraphics[width=3 in]{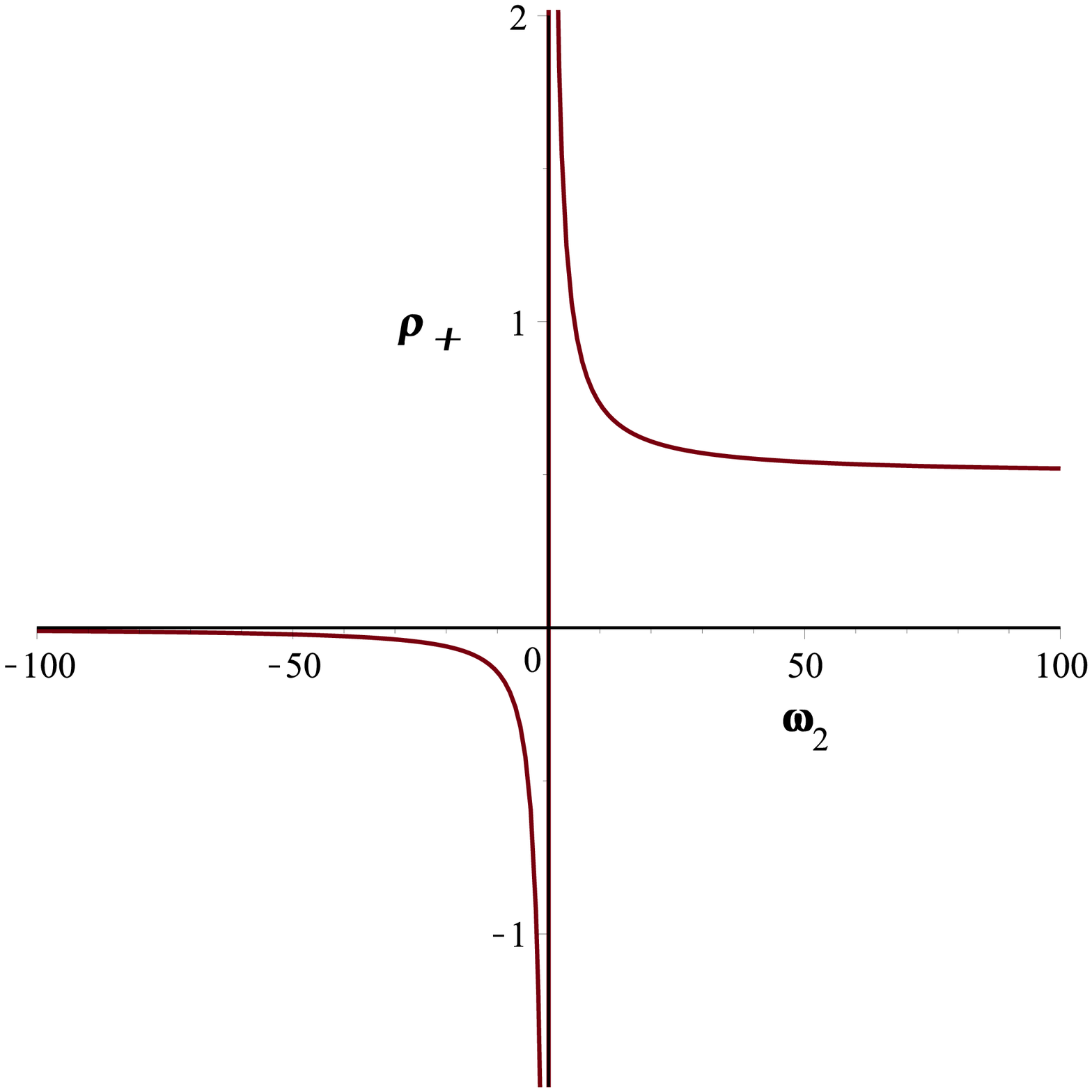}
\caption{The plot depicts the function $\rho_+(\omega_2)$ as a function of $\omega_2$ . It is clear that  $\rho_+$ is not in the possible range $0\leq \rho\leq \frac{1}{2}$ so $\rho_+$ is not  acceptable. See the tex for details.}
\label{fig9}
\end{figure}
\begin{figure}
\centering
  \includegraphics[width=3 in]{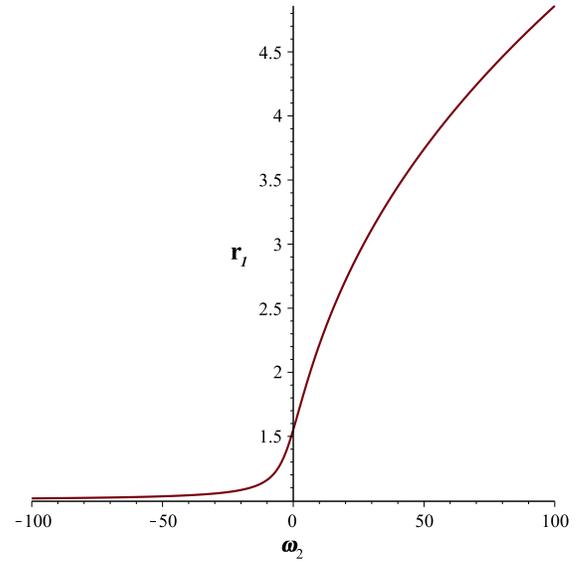}
\caption{The plot depicts the function $r_1(\omega_2)$ as a function of $\omega_2$ where $\omega=-\frac{1}{2}$. It is evident that  $r_1$ increases as $\omega_2$ increases which shows that the NEC violation is restricted in a smaller region for smaller $\omega_2$ . See the tex for details.}
\label{fig10}
\end{figure}
We have plotted $\rho_-(\omega_2)$ and $\rho_+(\omega_2)$ as a function of $\omega_2$ in Figs. \ref{fig8} and  \ref{fig9} respectively. Since for $\alpha=\frac{1}{2}$ , the possible range for energy density is $0\leq\rho\leq\frac{1}{2}$, one can deduce from these figures that  $\rho_+$ is not acceptable. Now, it is easy to show that
 \begin{eqnarray}\label{N9}
r_1(\omega_2)&=&(2\,\rho_-)^{-\frac{5}{2}}\nonumber  \\
&=&(-\frac{-3+\frac{\omega_2}{2}-\sqrt{(3+\frac{\omega_2}{2})^2-2\omega_2}}{\omega_2})^{-\frac{2}{5}}.
\end{eqnarray}
$r_1(\omega_2)$  has been plotted against $\omega_2$ in Fig. \ref{fig10}. This plot shows that $r_1$ is an ascending function of $\omega_2$. It means that the smaller $\omega_2$ will present a wormhole with smaller amount of the NEC violation.
 We analyze three cases $\omega_2=-10$ (case 1), $\omega_2=-2$ (case 2) and $\omega_2=0$ (case 3) which  are related to following EoS's respectively.
 \begin{eqnarray}\label{N88}
p_1&=&-\frac{1}{2}\rho+2\rho^{2}-10\rho^3, \nonumber  \\
p_2&=&-\frac{1}{2}\rho-2\rho^{2}-2\rho^3,
   \nonumber  \\
p_3&=&-\frac{1}{2}\rho-3\rho^{2}.
\end{eqnarray}

\begin{figure}
\centering
  \includegraphics[width=3 in]{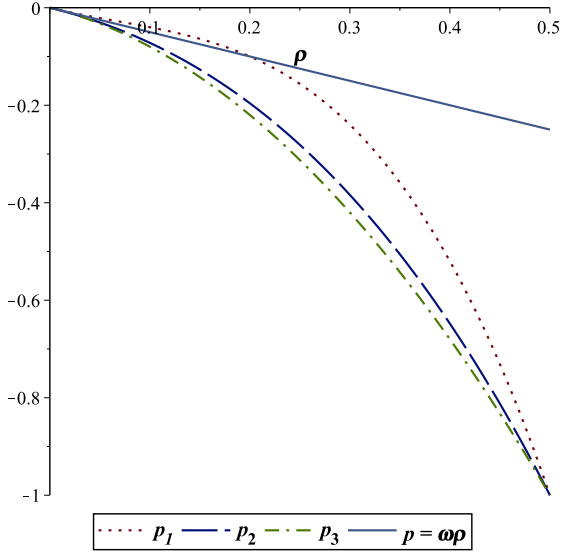}
\caption{The plot depicts the three special EoS $p_1$ (dotted line), $p_2$ (dashed line), $p_3$ (dotted-dashed line) and $p=\omega \rho$ (solid line) as a function of $\rho$. It is evident that  $p_1$ has the most consistency with   $p=\omega \rho$. So the EoS with extra terms can increase the opportunity  of consistency with linear EoS. See the tex for details.}
\label{fig11}
\end{figure}
We have plotted $p(\rho)$ as a function of $\rho$ for three cases and compared with $p=\omega\rho$ in Fig. \ref{fig11}. This figure shows that $p_1$ have more consistency with linear EoS. It should be noted from Fig. \ref{fig10} that the NEC is violated for $\omega_2=-10$ (case 1)   in a smaller region  in comparison to the other cases. It is also of particular interest that $p_3$ is a quadratic EoS which is achieved by putting $\omega_2=0$ in Eq.(\ref{N6a}). So one should note that if EoS has more  terms, the opportunity to  fit the EoS with linear EoS is more accessible  and the NEC violation is restricted to a smaller region. Also, EoS with few terms can be considered as a special case of a more general EoS. This result may be considered as the one of the main points of our work. It should be mentioned that the conditions (\ref{f1}) and (\ref{s1}) are necessary but not sufficient to check  wether wormhole is  traversable or not. We should checked the parameters
\begin{eqnarray}\label{N12}
|a(r)|&=&|(1-\frac{b(r)}{r})^{1/2}e^{-\phi(r)}(\gamma e^{\phi(r)})'| \\
|\bigtriangleup a_t(r)|&=&|(1-\frac{b(r)}{r})(-\phi''+\frac{b'r-b}{2r(r-b)}\phi'-(\phi')^2|
\end{eqnarray}
 where $a$ is the radial acceleration that traveler feels and $\bigtriangleup a_t$ is the maximum tidal acceleration felt by traveler \cite{WH}. As an example, for metric \ref{28a} we have calculated the maximum value of $|a(r)|$ in the range $1\leq r< \infty$ which is equal to $a_{max} = 0.52$. So, the condition
 \begin{equation}\label{N13}
|a(r)|\leq g_{\oplus},
\end{equation}
  is satisfied. Also, the maximum value of $\bigtriangleup a_t$ is $(\bigtriangleup a_t )_{max}=0.25$. Therefor, the condition
  \begin{equation}\label{N14}
|\bigtriangleup a_t(r)|\leq \frac{g_{\oplus}}{2}
\end{equation}
is satisfied. Note that we have considered $8\pi G=C=1$. In \cite{WH}, it was mentioned that
\begin{eqnarray}\label{N15}
\triangle \tau_1&=&\int_{-l_1}^{l_2}\frac{dl}{v\gamma} \leq 1yr \\
\triangle t_1 &=&\int_{-l_1}^{l_2} \frac{dl}{v e^{\phi}}\leq 1yr .
\end{eqnarray}
are the necessary conditions in a journey. Checking these conditions analytically for our solutions is so difficult but one can consider some value for $l_1$ and $l_2$ then try to check these conditions numerically. As an example, if we set $l_1=-l_2=-3\times 10^8$ for metric \ref{28a} then we can find $\triangle \tau_1\simeq19.89$ and $\triangle t_1\simeq20.002$ . So total time to travel conditions are satisfied. During this calculation, we consider $\gamma=1/\sqrt{0.99}$. Reader can used the same procedure for other solutions.


\section{Discussion and concluding remarks}\label{sec5}
Wormholes have attracted a lot of attention in the literature and a large number of models have been proposed in this realm. Many of these studies are presented to answer to the basic question, how to avoid or at least minimize the requirement of the exotic NEC violating matter?
Minimizing the usage of exotic matter for the physical viability of wormhole has received considerable attention.
We have shown the fact that the choice of EoS for the description of matter presenting  the wormhole has great relevance in the violation of the NEC in the wormhole static solutions. In the following discussion, we will see how our method is interpreted as a solution for  wormhole and how the definition of EoS   affects the violation of energy conditions. We also discuss the priority of this method than the cut and paste method.

Actually, we can divide the geometry of the wormhole into two regimes; high energy density regime and low energy density regime. If we consider $\rho$ as a steady function of $r$ with a minimum and a maximum, then low regime is related to the regime where $\rho$ tends to a minimum when  the high regime is relevance to the vicinity of the maximum.
This motivation helps us to consider an EoS with two terms. The first term is the dominant term in the low regime. The violation of the NEC can be caused by the combination of these two terms. On the other hand, it supports the possibility of confining the exotic matter in some  finite regions of the spacetime. The shape function (\ref{shape1}) present a smooth energy density function which has a maximum at the throat and tends to zero at large distance from the throat. It seems that this is a suitable choice to present the desired energy density. It should be noted that this algorithm can be used for other forms of shape functions or energy density. Any physical energy density which has a maximum near the throat and tend to zero at large distance may be considered as a candidate. The mathematical difficulty of solving equations  to find redshift function is different for different choices of shape function.

 Generally, in our method EoS is as follows
\begin{equation}\label{NN3}
p=L(\rho)+H(\rho).
\end{equation}
The former function($L$) in the recent equation gives information about the behavior of  the fluid supporting wormhole far from the throat while combination of ($L$)and ($H$) specifies the behavior  of the fluid in the vicinity of the wormhole throat. It should be noted that the portion of ($H$)is more than ($L$) near the wormhole throat. The parameter $\epsilon(r)=\frac{H}{L}$ can be used as a measure to quantify the portion of each term in different regions. It is clear that $\lim_{r\rightarrow \infty } \epsilon\rightarrow 0 $.
Since  $p(\rho)=\omega\rho$ is the most usual EoS which has been used in the literature to describe the cosmos, we set $L(\rho)=\omega\rho$ to have consistency with physics of cosmos in large scale. Note that Eq.(\ref{f4})provides the difference between the low regime and high regime in the EoS. This is the essential condition for the function $H(\rho)$. As an example, wormhole with a modified Chaplygin EoS \cite{jamil2}
\begin{equation}\label{NN4}
p=\omega \rho-\frac{\omega_1}{\rho^\alpha},
\end{equation}
with $\alpha>0$ is not a suitable case in our method. Because Eq.(\ref{f4}) is not satisfied.
 The use of the mixed EoS (\ref{NN3}) offers two important advantage over $p(\rho)=\omega\rho$: First, in the wormhole theory, violation of the NEC is an essential condition at the throat, so one should assume $\omega<-1$ for satisfying flaring out condition. This directed us toward phantom wormhole, but phantom models have problems such as Big Rip singularity or problem of instability in quantum field theory. In phantom model, the violation of the NEC is inevitable in the entire range of spacetime. Mathematically, this assumption is not necessary when the EoS is considered as a mixed model, thereby producing a smaller violation of energy conditions.
 Second, in the cut and paste method the violation of the NEC can be confined to a small region near the throat.
 The wormhole metric was matched to an exterior vacuum metric to keep the exotic matter within a finite region of space.

 We have found an asymptotically flat spacetime without having to go through acrobatics of cutting and
pasting. Our method seems to be more physically than this method. In the cut and paste method, the geometry is divided into two different parts the interior part is the wormhole line element and the exterior part is the geometry of the general cosmos spacetime or at least a geometry which asymptotically is compatible  with the metric of cosmos. The surgery may need a surface stress-energy tensor in the boundary of the cut and paste. In our method, the geometry is single and there is not any need to surface stress-energy tensor. The EoS is compatible with cosmos EoS in large scale intrinsically. We have utilized a specific shape function which describes a smooth energy density function with a maximum near the throat and tends to zero at large distance from the throat.  Although some classes of EoS are investigated in this paper, this method can be  used with the  other EoS or shape function to find new solutions. For solutions with quadratic EoS, avoiding horizon at the throat has led to a relation between $\omega$ and $\omega_1$. The same condition is imposed between the $\omega_i$ parameters  on the other EoS. So we can not choose EoS without considering this constraint. We have shown that for a combination of barotropic and  polytropic EoS $r_1$ is dependent on the parameter $\omega, n$  and $\alpha$ in which the larger $n$ proposed a fluid with a smaller amount of exotic matter. The general dependence  of  $r_1$  has been discussed in details. We have presented solutions with a more general EoS which are a polynomial function of $\rho$. The previous  solutions can be considered as a special case of these solutions. We have shown that for three individual cases (Eq.(\ref{N88})), if we choose an EoS with more additional terms, the opportunity to have more consistency with linear EoS will increase. Also, one can conclude that a more general EoS instead of a linear one may present solutions with fewer violation of the NEC. As we know, different  astrophysical objects may have different EoS in the local view in comparison with  global view.
 In the other word, the EoS of fluid supporting wormhole near the throat has a behavior different from the behavior of  the EoS of the Universe in large scale. This local behaviour of fluid near the throat will tend to a global EoS in the large scale. This method seems to be more physically than considering a unique EoS in all regions of the spacetime. This is the main point of our work which leads to the minimum violation of the NEC.

\section*{Acknowledgements}
This work has been in part supported by a grant from the Research Council of Sirjan University of Technology.

\end{document}